\documentclass[a4paper]{jpconf}
\usepackage{graphicx, caption}
\usepackage{hyperref}
\usepackage[frozencache=true,cachedir=minted-cache]{minted}
\usepackage{amsmath}

\begin{document}
\pagenumbering{arabic}
\title[Automatic Differentiation of Binned Likelihoods With Roofit and Clad]{Automatic Differentiation of Binned Likelihoods With Roofit and Clad}

\author{Garima Singh\textsuperscript{*}, 
Jonas Rembser\textsuperscript{$\dag$}, 
Lorenzo Moneta\textsuperscript{$\dag$},
David Lange\textsuperscript{*}, 
Vassil Vassilev\textsuperscript{*}}

\address{* Department of Physics, Princeton University, Princeton, NJ 08544, USA}

\address{$\dag$ EP-SFT, CERN, Espl. des Particules 1, 1211 Meyrin, Switzerland}

\ead{\mailto{garima.singh@cern.ch}, \mailto{jonas.rembser@cern.ch}, \mailto{lorenzo.moneta@cern.ch}, \mailto{david.lange@cern.ch}, \mailto{vassil.vassilev@cern.ch}}

\begin{abstract} Just as data sets from next-generation experiments grow, processing requirements for physics analysis become more computationally demanding, necessitating performance optimizations for RooFit. One possibility to speed-up minimization and add stability is the use of Automatic Differentiation (AD). Unlike for numerical differentiation, the computation cost scales linearly with the number of parameters, making AD particularly appealing for statistical models with many parameters. In this paper, we report on one possible way to implement AD in RooFit. Our approach is to add a facility to generate C++ code for a full RooFit model automatically. Unlike the original RooFit model, this generated code is free of virtual function calls and other RooFit-specific overhead. In particular, this code is then used to produce the gradient automatically with Clad. Clad is a source transformation AD tool implemented as a plugin to the clang compiler, which automatically generates the derivative code for input C++ functions. We show results demonstrating the improvements observed when applying this code generation strategy to HistFactory and other commonly used RooFit models. HistFactory is the subcomponent of RooFit that implements binned likelihood models with probability densities based on histogram templates. These models frequently have a very large number of free parameters and are thus an interesting first target for AD support in RooFit.
\end{abstract}

\section{Introduction}

The complexity of analyses of data collected at the Large Hadron Collider is growing continuously.
Cutting-edge statistical analyses such as combined Higgs measurements are usually executed with the RooFit library~\cite{Verkerke:2003ir}, a part of the data analysis framework ROOT~\cite{rene_brun_2019_3895860}.
In these studies, one has to minimize likelihoods with thousands of free parameters that spread over hundreds of likelihood components, each representing a different measurement channel.
For the largest models, the likelihood minimization can take hours to complete~\cite{zeffichep}.
To ensure that this time doesn't grow further and hopefully even bring it down to the order of minutes, RooFit's performance was significantly improved in the last years. This has included, for example, a complete rewrite of the likelihood evaluation code to achieve easier vectorization~\cite{Michalainas:2023rtu}.

To iteratively find the minimum of a likelihood, one generally has to know the gradient of the likelihood with respect to all free parameters.
Naively, the gradient can be found numerically by varying one parameter at a time and reevaluating the full likelihood.
However, it is not necessary to reevaluate the full mathematical expression when only one parameter is changed, which is why RooFit includes a sophisticated caching mechanism. Individual expressions in the computation graph or groups thereof are represented by objects that cache their last results.

This caching approach has some performance shortcomings. It requires lots of bookkeeping (such as keeping track of the parts of the model that need to be recomputed due to a change in the parameter values) that incurs significant overhead.
The representation of the computation graph in terms of objects with a relatively large memory footprint that evaluates each other using virtual calls is also not ideal for performance.
Furthermore, there are still some redundant computations because a single RooFit object often groups many mathematical operations together.
In other words, for some subgraphs of the computation graph, we fall back to the fully redundant reevaluation.

Deviating from this caching paradigm, we believe that all these performance penalties can be alleviated by better analyzing the full expression graph for the likelihood at the most granular level to automatically generate an overhead-free representation for the full gradient without redundant computations.
This is a job that can be done by the compiler.
For this paper, we generated the gradient for binned HistFactory-style likelihoods~\cite{Cranmer:1456844} with Clad~\cite{Vassilev_Clad}, a compiler-based source code transformation AD tool.
\section{Background}

\subsection{Numerical Differentiation}

Numerical differentiation aims to estimate the derivative of a mathematical function, specifically using the values of the function or other meta-information. It is widely used in popular mathematical software and libraries, including RooFit, where it serves as the main source of derivative calculation through the minimization library - MINUIT~\cite{James:1975dr}.

Numerical differentiation can be implemented through many different techniques, one of the most notable being Newton's method of finite difference. Here, the derivative of a real-valued multi-argument function \(f(x_0, x_1, ..., x_n)\) with respect to the input \(x_i\) can be written as follows:

\begin{equation}
\centering
\frac{\delta f(x_0, x_1, ..., x_n)}{\delta x_i} \equiv \frac{f(x_0, ..., x_i + h, ..., x_n) - f(x_0, ..., x_i - h, ..., x_n)}{2h}
\end{equation}

The method of finite differences relies on introducing a small perturbation \(h\) to the function and measuring its impact on the result of the function. This small perturbation is usually referred to as the \textit{step size}. The smaller the step size, the more accurate the derivative estimate. In theory, the step size should be a very small value; however, due to the nature of floating-point arithmetic, division by small floating-point values can lead to high numerical error in the result. This leads to a fairly difficult optimization problem for calculating the value of \(h\), as too small a value will lead to high numerical error, but too large a value will lead to worse estimates of the derivative. Moreover, this choice also depends on the problem being solved, meaning that there is no single value of \(h\) that is suitable for all use cases. As such, choosing the right step size while maintaining the accuracy of the derivative becomes very challenging.

Another aspect where numerical differentiation falters is performance. For example, in equation 1, the full derivative with respect to all inputs of the function \(f\) is calculated one at a time. This means that since each derivative calculation takes at least 2 calls (even higher for other more accurate methods), a function with \(n\) parameters will require at least \(2*n\) function calls to \(f\) to generate a full derivative. As such, these methods tend to scale very poorly when applied to problems with many parameters. However, one advantage of using numerical differentiation is the fact that minimal effort is needed to support complex functions and parameters. As shown in equation 1, most numerical differentiation techniques are fairly implementation agnostic, meaning they can handle arbitrarily complex function implementations as long as the interface of the function follows the one dictated by the numerical differentiation algorithm. For some uses, this advantage can trump some of the previously stated problems.

RooFit uses numerical differentiation for this exact reason; it is fairly simple to implement even for complex codebases. As for the other disadvantages, RooFit combats them by using sophisticated caching mechanisms and other numerical logic that reduces the final error in the derivative while still keeping the performance satisfactory. These modifications have led to RooFit possessing an efficient numerical differentiation pipeline that can handle large-scale analysis effectively at the cost of making the codebase very complex, hindering the maintenance and development of new features.   

\subsection{Automatic Differentiation and Clad}
Automatic Differentiation (AD) is a set of techniques to evaluate the exact derivative of a computer program. Unlike numerical differentiation, AD can produce true derivatives rather than estimates. It uses the  chain rule of differentiation over a function's computation graph to differentiate it.

AD majorly defines two types of recursive relationships to calculate the derivative of a function \(f\) - forward accumulation and reverse accumulation. For an arbitrarily complex function \( y = f(g(x))\) that can be split into intermediate steps as follows:

\begin{equation}
  w_0 = x
  \quad
  w_1 = g(x)
  \quad
  w_2 = f(g(x)) = y
\end{equation}

The forward and reverse accumulation modes are defined as follows: 

\begin{equation}
\frac{\partial{w_i}}{\partial{x}} = \frac{\partial{w_i}}{\partial{w_{i-1}}} * \frac{\partial{w_{i-1}}}{\partial{x}}
   \quad\mathrm{and}\quad 
\frac{\partial{y}}{\partial{w_i}} = \frac{\partial{y}}{\partial{w_{i+1}}} * \frac{\partial{w_{i+1}}}{\partial{w_i}}
\end{equation}

\begin{figure}[H]
    \centering
    \includegraphics[width=\linewidth]{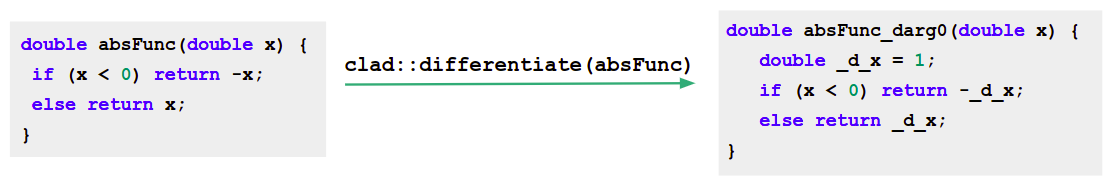}
    \caption{\textbf{An example function being differentiated by Clad.}}
    \label{fig:cladfunc}
    \vspace{-2mm}
\end{figure}

As described above, the forward accumulation mode calculates the derivative of each sub-expression in a function with respect to a fixed input (much like numerical differentiation), whereas the reverse accumulation mode calculates the derivative of the output w.r.t. to each sub-expression. This quality of the reverse mode makes it an ideal choice for when the function needs to be differentiated w.r.t. to a large number of parameters, as it entails lesser work than the forward accumulation mode. For this reason, we primarily use the reverse accumulation mode (or \textit{reverse mode}) in our work. Specifically, we use Clad~\cite{Vassilev_Clad}, a source-code-transformation AD tool built as a plugin for the C/C++ Clang compiler based on the LLVM compiler infrastructure~\cite{LLVM:CGO04}. Clad works at the compiler level to inspect the source code of a target function and generates another function that calculates its derivative. Figure~\ref{fig:cladfunc} demonstrates what an example function and its derivative look like when differentiated with Clad. Clad is especially easy to use for HEP tools as it is available in ROOT through Cling~\cite{Cling}, the C++ interpreter for ROOT.

While AD has clear advantages of efficiency over numerical differentiation, one caveat it has is its limited support for code with side effects (such as classes and other object-oriented programming constructs). In a lot of cases, these constructs can hide the vital differentiable information that AD needs to generate correct derivatives. While this 'hiding' is intentional to make the target application easy to use, it becomes a hindrance for AD tools. This makes AD particularly difficult to integrate into RooFit, where classes and object-oriented programming are important design elements.    

\section{Design and Implementation}

As the previous sections discussed, RooFit's object-oriented design makes it hard to directly use an AD tool to differentiate the statistical models. We need to introduce an intermediary step or transformation that can make these models more amenable to AD. Using a tool like Clad, this means generating state-independent code that exposes the differentiable properties of a RooFit model in a way that Clad can understand. 

\begin{figure}
    \centering
    \includegraphics[width=\linewidth]{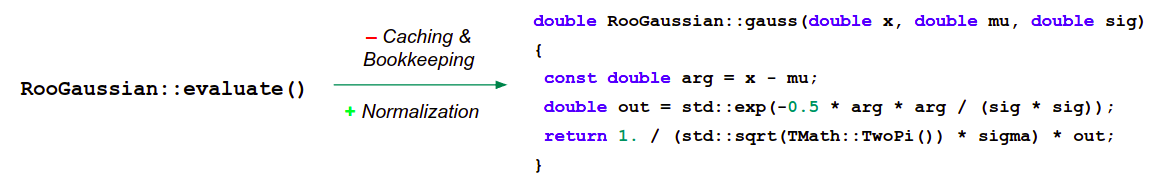}    \caption{\textbf{An example showing how a RooFit class can be transformed to expose its differential properties.}}
    \label{fig:codetrans}
    \vspace{-3mm}
\end{figure}

The first step in building such a transformation is exposing the relevant differential properties of the RooFit classes. We can accomplish it by introducing a static function that translates the mathematical meaning of the class into code. Fig.~\ref{fig:codetrans} shows how the \mintinline{cpp}{RooGaussian::evaluate()} can be transformed to create a function that can be differentiated by Clad.

Now that we can transform each class independently, we need to combine all of these transformations from the statistical model into a single piece of code that can later be differentiated by Clad. While it is possible to combine the independent derivatives from each of the nodes/classes in a model to get the derivative of the full model, it usually is not as efficient. One big advantage of first combining the code and then generating the derivative is that the compiler has better opportunities to perform compiler optimizations on both the combined code and generated derivative. Moreover, this approach of \textit{combine-and-differentiate} also leads to a better debugging experience as the cause of errors in the derivative can be attributed directly to a line of stateless code rather than a deep stack trace of member functions. For these reasons, in this work, we use the \textit{combine-and-differentiate} approach.

One way to implement the \textit{combine-and-differentiate} (or \textit{code-squashing}) approach is by introducing a "translate" function for each of the classes. This "translate" function returns an \mintinline{cpp}{std::string} representing the underlying mathematical notation of the class as code that can later be concatenated into a single string representing the entire model.
\begin{figure}[H]
    \centering
    \includegraphics[width=\linewidth]{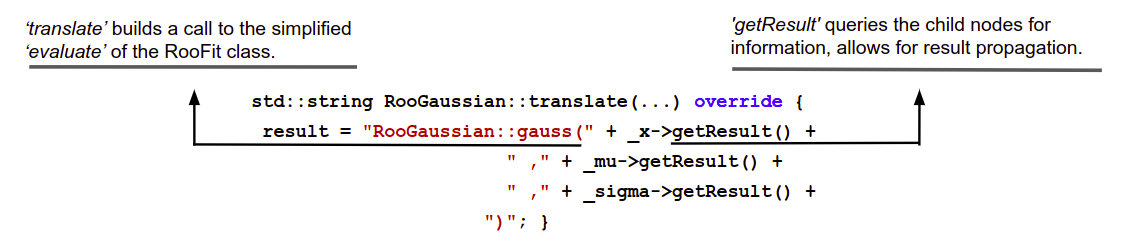}
    \caption{\textbf{An example "translate" function for \mintinline{cpp}{RooGaussian}.}}
    \label{fig:codetransfunc}
    \vspace{-3mm}
\end{figure}
Moreover, it also contains some extra logic to correctly propagate the required information (such as inputs to other nodes, side effects of the current node on the other nodes, etc.) to the children nodes. An example of a \mintinline{cpp}{translate} function for the \mintinline{cpp}{RooGaussian} class is given in fig.~\ref{fig:codetransfunc}. 

\begin{figure}[h]
    \centering
    \includegraphics[width=\linewidth]{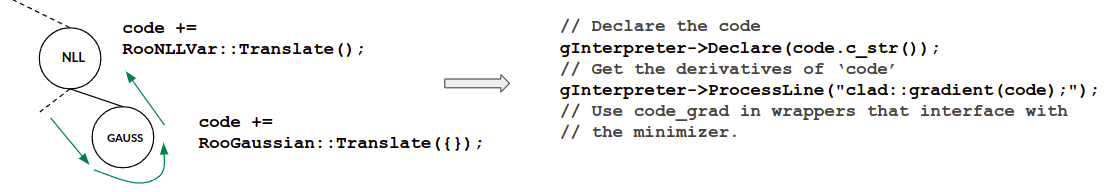}
    \caption{\textbf{Code generation and differentiation for an example RooFit model.}}
    \label{fig:codecompl}
    \vspace{-4mm}
\end{figure}

To retain the control flow, a depth-first search is performed on the model, and \mintinline{cpp}{translate} is called on each of the nodes being visited. Once the search is complete, the generated code is just-in-time compiled using Cling, the C++ interpreter for ROOT. Fig.~\ref{fig:codecompl} describes how an example RooFit model can be translated into code and then differentiated by Clad.

\section{Results}

We tested our work with an example based on the HistFactory template \cite{Cranmer:1456844}. The negative log-likelihood function looks as follows:
\begin{equation}
l = \sum_i \text{pois}(n_i|\alpha \cdot \nu_i) + l_{\text{aux}}.
\label{eqn:model}
\end{equation}

Here, $\text{pois}(n_i|\alpha \cdot \nu_i)$ (in equation~\ref{eqn:model}) denotes the negative logarithm of a Poisson distribution with the expected value $\alpha \cdot \nu_i$ at the observed value $n_i$ in the bin $i$. The expected value is the product of a global luminosity scaling constant $\alpha$ and the sum of three scaled template histograms $S$, $B_1$ and $B_2$, standing for one signal sample and two background samples:
\begin{equation}
\nu_i = \mu S_i + \gamma_{1,i}B_{1,i} + \gamma_{1,2}B_{2,i}.
\label{eqn:model2}
\end{equation}
The scaling factor $\mu$ (in equation~\ref{eqn:model2}) for the signal is independent of the bin, and the $\gamma$-factors are independent for each bin and background. The parameter $\alpha$ is constrained by a Gaussian distribution, while each $\gamma$ is constrained by a Poisson distribution.
All these constraints are included in the auxiliary likelihood $l_\text{aux}$. This is typical for measuring a predicted signal on top of data-driven background estimates. For our study, we varied the number of parameters in the model by varying the number of bins.
\begin{figure}[h]
    \centering
 \includegraphics[width=\linewidth]{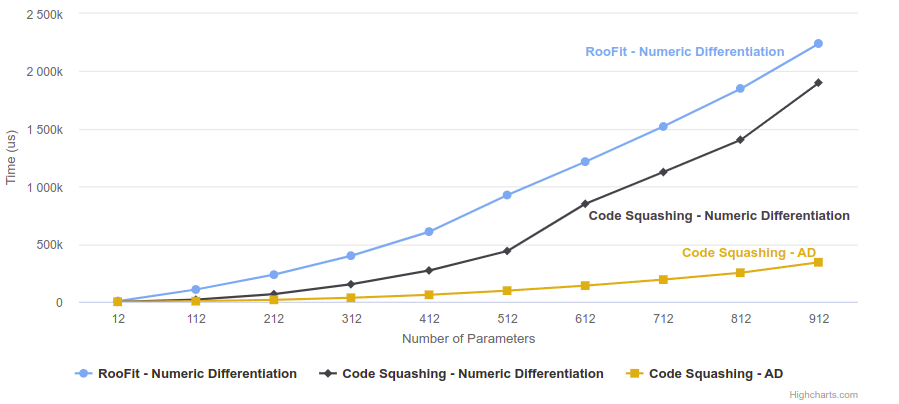}
    \caption{\textbf{Performance comparison AD vs numerical differentiation on a HistFactory template (defined in equation~\ref{eqn:model}).}}
    \label{fig:finGraph}
    \vspace{-4mm}
\end{figure}

Fig~\ref{fig:finGraph} shows the performance comparison of the minimization of the model defined in equation~\ref{eqn:model} for different evaluation configurations. We compared the combined code (or \textit{code-squashing}) approach using AD and numerical differentiation with the regular RooFit configuration for a varied number of parameters. We obtain a speedup of \textbf{5.5x} for the minimization with around 1,000 parameters.
\section{Conclusion and Future Work}

This work has shown that RooFit's minimization performance can be significantly improved by generating an optimized callable for the analytic gradient instead of using numeric differentiation with caching in an attempt to avoid redundant computations.
The future work will aim to implement this logic upstream in ROOT such that it can be used for general RooFit models. To this end, future work is needed in three areas. The first area is the implementation of a code generation engine that can take any RooFit likelihood and produce a simple C++ function from it that can be digested with Clad.
The second area is to improve the interface between RooFit and the minimizer algorithms shipped with ROOT, as the many (often unnecessary) abstraction layers make passing around the analytic gradient very complicated.
The third area is to work on the minimization algorithms themselves.
Having a callable representation for the first and potentially also second-order derivatives of the likelihood that are cheaper to evaluate than the numeric approximations makes it necessary to revisit the minimization strategies in order to maximally benefit from the analytic gradients and Hessians.
\section{Acknowledgements}

 This project is
supported by the National Science Foundation under Grant OAC-1931408 and under Cooperative Agreement OAC-1836650.
\section{References}

\bibliographystyle{IEEEtran}
\bibliography{main}
\end{document}